%% file: WASP-12b.tex
\documentclass[twocolumn,appendixfloats]{aastex62}

\input macros.tex

\received{25 Jul 2018}
\revised{19 Aug 2018}
\accepted{21 Aug 2018}
\submitjournal{The Astronomical Journal}

\usepackage{graphicx}
\usepackage{natbib}
\usepackage{latexsym}
\usepackage{amssymb}
\usepackage{longtable}
\usepackage{amsmath}
\usepackage{url}
\usepackage{lipsum}
\usepackage{graphicx}

\def\msun{\ifmmode {\rm\,M_\odot}\else ${\rm\,M_\odot}$\fi}
\def\Msun{\ifmmode {\rm\,\it{M_\odot}}\else ${\rm\,M_\odot}$\fi}
\def\lsun{\ifmmode {\rm\,L_\odot}\else ${\rm\,L_\odot}$\fi}
\def\Lsun{\ifmmode {\rm\,\it{L_\odot}}\else ${\rm\,L_\odot}$\fi}
\def\rsun{\ifmmode {\rm\,R_\odot}\else ${\rm\,R_\odot}$\fi}
\def\Rsun{\ifmmode {\rm\,\it{R_\odot}}\else ${\rm\,R_\odot}$\fi}
\def\Tsun{\ifmmode {\rm\,T_\odot}\else ${\rm\,T_\odot}$\fi}
\def\arcsec{\ifmmode {^{\prime\prime}}\else $^{\prime\prime}$\fi}
\def\asec{\ifmmode {^{\prime\prime}}\else $^{\prime\prime}$\fi}
\def\arcmin{\ifmmode {^{\prime}}\else $^{\prime}$\fi}
\def\amin{\ifmmode {^{\prime}}\else $^{\prime}$\fi}
\def\simlt{\mathrel{\spose{\lower 3pt\hbox{$\mathchar"218$}}
     \raise 2.0pt\hbox{$\mathchar"13C$}}}
\def\simgt{\mathrel{\spose{\lower 3pt\hbox{$\mathchar"218$}}
\     \raise 2.0pt\hbox{$\mathchar"13E$}}}

\def\fmeas{F_{\rm meas}}
\def\fref{F_{\rm ref}}

\shorttitle{WASP-12b Transmission Spectra}

\shortauthors{Jensen et al.}

\begin{document}

\title{HYDROGEN AND SODIUM ABSORPTION IN THE OPTICAL TRANSMISSION SPECTRUM OF WASP-12b}



\author{Adam G.~Jensen}\affiliation{Department of Physics and Astronomy, University of Nebraska at Kearney, 2401 11th Ave, Kearney, NE 68849; JensenAG@unk.edu}
\author{P.~Wilson Cauley}\affiliation{School of Earth and Space Sciences, Arizona State University, Tempe, AZ 85287; pwcauley@gmail.com}\altaffiliation{Current affiliation:  \\ Center for Astrophysics and Space Astronomy \\ University of Colorado at Boulder \\ 389 UCB, Boulder, CO 80309}
\author{Seth Redfield}\affiliation{Van Vleck Observatory, Astronomy Department, Wesleyan University, 96 Foss Hill Drive, Middletown, CT 06459; sredfield@wesleyan.edu}
\author{William D.~Cochran}
\author{Michael Endl}\affiliation{Department of Astronomy, University of Texas, Austin, TX 78712; wdc@astro.as.utexas.edu, mike@astro.as.utexas.edu}

\correspondingauthor{Adam G.~Jensen}
\email{JensenAG@unk.edu}

\begin{abstract}
We have obtained $> 10$ hours of medium resolution ($R \sim 15000$) spectroscopic exposures on the transiting exoplanet host star WASP-12, including $\sim2$ hours while its planet, WASP-12b, is in transit, with the Hobby-Eberly Telescope (HET).  The out-of-transit and in-transit spectra are coadded into master out-of-transit and in-transit spectra, from which we create a master transmission spectrum.  Strong, statistically significant absorption features are seen in the transmission spectrum at H$\alpha$ and \ion{Na}{1} (the Na D doublet).  There is the suggestion of pre- and post-transit absorption in both H$\alpha$ and \ion{Na}{1} when the transmission spectrum is examined as a function of phase.  The timing of the pre-transit absorption is roughly consistent with previous results for metal absorption in WASP-12b, and the level of the \ion{Na}{1} absorption is consistent with a previous tentative detection.  No absorption is seen in the control line of \ion{Ca}{1} at $\lambda$6122.  We discuss in particular whether or not the WASP-12b H$\alpha$ absorption signal is of circumplanetary origin---an interpretation that is bolstered by the pre- and post-transit evidence---which would make it one of only a small number of detections of circumplanetary H$\alpha$ absorption in an exoplanet to date, the most well-studied being HD 189733b.  We further discuss the notable differences between the HD 189733 and WASP-12 systems, and the implications for a physical understanding of the origin of the absorption.
\end{abstract}

\section{INTRODUCTION AND BACKGROUND}
\label{s:intro}

\subsection{The Hot Jupiter WASP-12b}
\label{ss:wasp12b}
WASP-12b is a very short-period hot Jupiter \citep{Hebb2009} discovered with the SuperWASP program \citep{Pollacco2006}.  At the time of its discovery it was the most highly-irradiated and shortest period planet in the literature, and has been the subject of significant attention ever since.  Shortly after its discovery, \citet{Fossati2010} found evidence suggestive of absorption by \ion{Mg}{2} and other metals in WASP-12b's atmosphere through {\it HST}/COS observations.  These results include an indication of pre-transit absorption, potentially signifying that WASP-12b has a significant extended atmosphere.  \citet{Haswell2012} presented a second {\it HST}/COS visit that bolstered the results of \citet{Fossati2010}, including a detection of \ion{Fe}{2}.  The \citet{Haswell2012} results showed that the required interpretation of the system was more complex, especially given that the ingress of the second visit began significantly earlier than the first visit in \citet{Fossati2010}.  The depth of the transits indicate that the planet's Roche lobe is overfilled \citep{Lai2010,Debrecht2018}, and there is also observational and theoretical evidence that material from WASP-12b forms a torus- or disk-like structure around the star \citep{Haswell2012,Fossati2013,Debrecht2018}.

The C/O ratio and water content of WASP-12b have also been the focus of much study, especially the possibility that the planet is carbon-rich rather than oxygen-rich.  Evidence of a high C/O ratio and a weak thermal inversion was found by \citet{Madhusudhan2011}.  More recently, \citet{Kreidberg2015} found evidence for water in the HST/WFC3 transmission spectrum of WASP-12b.  This detection is potentially consistent with high C/O ratios ($>1$) under certain assumptions, but strongly favors a C/O ratio closer to 0.5.

Another question of interest is whether WASP-12b has an absorbing layer of TiO/VO.  The nominal expectation from using the classification system of \citet{Fortney2008} is that WASP-12b should have TiO/VO in its atmosphere that masks \ion{Na}{1} absorption.  However, \citet{Sing2013} found evidence of a lack of TiO in WASP-12b in {\it HST} observations.  Recently, \citet{Burton2015} tentatively detected \ion{Na}{1} in WASP-12b's atmosphere through defocused transmission spectroscopy at the level of $0.12\pm0.03[\pm0.3]$\%, with the error in brackets representing additional systematic uncertainty; accounting for the uncertainty they state that $0.15$\% is a better representation of their absorption measurement.  This detection of \ion{Na}{1} is consistent with the \citet{Sing2013} results and further indicates that WASP-12b does not fit neatly into the \citet{Fortney2008} classification system.  Furthermore, \citet{Sing2013} find that their transmission spectrum can be fit equally well by either Rayleigh or Mie scattering.  However, Mie scattering is a much better fit to the expected atmospheric temperatures (and the observed blackbody emission spectrum), and implies a high-altitude haze.  This subsequently means that any observed line absorption must occur at high altitudes above the haze.

One final recent WASP-12b observation is relevant to the current work, a search for \ion{He}{1} absorption at 10833 \AA{} by \citet{KriedbergOklopcic2018}.  They found a transit depth of $59\pm143\ppm$ relative to the adjacent wavelength bands in Hubble Space Telescope/Wide Field Camera 3 G102 grism data.  This non-detection does constrain certain models for WASP-12b's atmosphere and indicates that any helium absorption in WASP-12b is smaller than a recent detection in WASP-107b \citep{Spake2018}.  However, the measurement is made over an integration band of 70 \AA{} (the instrumental resolution) and does not rule out the possibility that significant absorption might be observed with a higher instrumental resolution.

\subsection{Observations of Hydrogen in Exoplanets}
\label{ss:hinexo}
Observations of broad hydrogen envelopes measured in ground state absorption (specifically in Ly$\alpha$) have been made in multiple planets, including the well-studied planets HD 209458b {\citep{VidalMadjar2003,VidalMadjar2004} and HD 189733b \citep{LecavelierDesEtangs2010,LecavelierDesEtangs2012}.  Such observations give insight into the possibility of star-planet interactions and atmospheric escape.  However, exoplanetary Ly$\alpha$ measurements pose significant observational difficulties.  First, Ly$\alpha$ is in the UV, and UV-capable facilities with adequate throughput and spectral resolution are a very limited resource; only {\it HST} is capable of such observations in Ly$\alpha$ at the current time.  Second, certain stars are not bright at Ly$\alpha$, especially sun-like stars.  Furthermore, interstellar Ly$\alpha$ absorption is significant, and absorbs stellar Ly$\alpha$; this absorption poses difficulties for all observations of exoplanetary Ly$\alpha$ and makes them completely impractical in lines of sight longer than 50 pc or so.  In short, only certain relatively nearby systems will have adequate observable Ly$\alpha$ flux available to perform exoplanetary transmission spectroscopy.  The aforementioned detection of \ion{He}{1} at 10833 \AA{} in WASP-107b \citep{Spake2018} is one promising observational strategy for observing extended exoplanetary atmospheres without UV facilities.  Another option is to directly observe hydrogen, albeit not in the ground state, through its Balmer series of transitions at visible wavelengths.

In a series of papers \citep[][hereafter Papers I, II, and III]{Redfield2008, Jensen2011, Jensen2012} the current authors examined the Hobby-Eberly Telescope (HET) transmission spectra of four hot Jupiter-like planets for absorption due to \ion{Na}{1}, \ion{K}{1}, and H$\alpha$.  Paper I made the first ground-based observation of an exoplanetary atmosphere by detecting \ion{Na}{1} in HD 189733b.  Paper II confirmed previously observed atmospheric \ion{Na}{1} absorption in HD 189733b (Paper I) and HD 209458b \citep{Charbonneau2002}, and found the hint of possible \ion{Na}{1} absorption in HD 149026b.  In Paper III, H$\alpha$ was detected in HD 189733b's transmission spectrum, the first-ever such detection of exoplanetary H$\alpha$.  Because the HET observations do not lend themselves to complete light curves that encompass an entire transit, observations with Keck I/HIRESr were obtained to observe a full transit of HD 189733b.  The first observations, in 2013, confirmed the H$\alpha$ transit absorption and found corresponding H$\beta$ and H$\gamma$ transit absorption, along with a significant pre-transit signal in all three lines \citep{Cauley2015}.  Subsequent observations in 2015 found strong variation in the pre-transit and transit signals, leading to an uncertain physical model for the geometry of the absorption \citep{Cauley2016}.

\citet{Barnes2016} challenged the circumplanetary interpretation of the H$\alpha$ absorption observed in HD 189733b, presenting an alternate interpretation that the transmission spectrum is dominated by contrast effects during transit.  Any star's spectrum will vary over its disk (due to spots or other active regions), and it is possible that a transiting planetary disk may result in differential spectroscopy that mimics excess absorption in certain lines.  However, there remain two significant arguments for a circumplanetary origin of the H$\alpha$ absorption in HD 189733b.  First, short-cadence monitoring of HD 189733's stellar H$\alpha$ shows that variation at the level of the observed pre-transit signals (which cannot be explained by the contrast effect) is uncommon \citep{Cauley2017b}.  Second, modeling of the stellar contrast effect that might mimic absorption during transit demonstrates that creating the absorption signal seen in HD 189733b requires a highly constrained distribution of active regions distributed only along the transit chord and/or a potentially implausible level of stellar activity for HD 189733 \citep{Cauley2017c}.

As of early 2018, HD 189733b was the only exoplanet with a published detection of H$\alpha$ absorption.  However, there are two recent detections of H$\alpha$ in exoplanetary atmospheres, both in planets around A stars---KELT-9b \citep{Yan2018} and MASCARA-2b/KELT-20b \citep{Casasayas2018}.  \citet{Yan2018} detected 1.15\% of extra absorption at H$\alpha$ line center in KELT-9b, and attribute the H$\alpha$ absorption to a hot extended atmosphere that is driven by the intense UV radiation of the star.  In addition to finding H$\alpha$, \citet{Casasayas2018} detected \ion{Na}{1} in KELT-20b.  They found that a temperature higher than equilibrium (4210 K vs.~2260 K, respectively) is required in order to explain the observations, which could be explained by the large amount of UV energy delivered by the central star.  

\subsection{WASP-12b and Comparative Planetology}
\label{ss:planetology}
In this paper we present the HET transmission spectrum of WASP-12b. This target was selected for HET observations as a point of comparison to the Paper III HD 189733b H$\alpha$ detection \citep[and prior to the Keck follow-up in][]{Cauley2015, Cauley2016}.  The central star WASP-12 has been classified as a G0V star by \citet{Bergfors2013}; however, \citet{Fossati2010star} find a temperature of $6250\pm100$ K, consistent with the value \citet{Hebb2009} derive, which would suggest a spectral type of approximately F7.  In any case, as it pertains to the possibility of H$\alpha$ absorption, WASP-12b is closer to a hotter central star than HD 189733b, and thus more highly irradiated.  On the other hand, HD 189733 is a later-type (K0V) star that is presumably more active than WASP-12, and its Ly$\alpha$ emission is likely to be both stronger and more variable.  This is significant as \citet{Huang2017} modeled the H$\alpha$ emission in HD 189733b, and found that stellar Ly$\alpha$ emission and Lyman continuum emission play an important role in creating the $n=2$ hydrogen population, although collisional excitation also plays a role \citep{Christie2013}.  Furthermore, \citet{Huang2017} suggest that HD 189733's activity may be the cause of the variation in the H$\alpha$ transit depth between \citet{Cauley2015} and \citet{Cauley2016}.  Thus, any H$\alpha$ detected in WASP-12b, given its intermediate spectral type and different physical conditions, would immediately be an interesting data point to enable comparative planetology for any exoplanets exhibiting transit-correlated H$\alpha$ absorption, including HD 189733b, KELT-9b, and KELT-20b.  Table \ref{table:parameters} provides information on the WASP-12 system.

\begin{deluxetable*}{ccc}
\tablecolumns{3}
\tabletypesize{\small}
\tablecaption{WASP-12 System Parameters\tablenotemark{a}\label{table:parameters}}
\tablehead{\colhead{Parameter} & \colhead{WASP-12/WASP-12b} & \colhead{Unit}}
\startdata
Transit Midpoint & $2454508.97682\pm0.0002$ &  HJD \\
Period & $1.09142245\pm3\times10^{-7}$ & days \\
Transit Duration & $0.122\pm0.001$ & days \\
$R_P$ & $1.79\pm0.09$ & $R_{\rm Jupiter}$ \\
$R_\star$ & $1.63\pm0.08$ & $R_\odot$ \\
$a/R_\star$ & $2.98\pm0.154$ & N/A \\
$b$ & $0.375^{+0.042}_{-0.049}$ & N/A \\
$i$ & $82.5^{+0.8}_{-0.7}$ & degrees \\
\enddata
\tablenotetext{\rm a}{All values from \citet{Maciejewski2011} and references therein, including \citet{Hebb2009}.}
\end{deluxetable*}

In \S\ref{s:obsdata} we describe our observations and data reduction, and in \S\ref{s:analysis} we describe our analysis methods.  Our results are presented in \S\ref{s:results}.  We discuss our results and possible future work in \S\ref{s:conclusions}.

\section{OBSERVATIONS AND DATA REDUCTION}
\label{s:obsdata}
\subsection{Description of the Observations}
\label{ss:obsdescription}
Observations of the WASP-12 system were obtained with the HET in March and April of 2012.  Spectra were observed with the High-Resolution Spectrograph (HRS) at its lowest resolution setting of $R\sim15000$ with a $2\arcsec$ fiber.  This is a lower resolution than used in Papers I--III, due to the relative faintness of WASP-12 ($V=11.7$) as compared to the targets in those papers.  Exposures for WASP-12 were 600 s in length.  The star HR 2866 was used as a telluric standard star; exposures of 60 s were used for this target.  HR 2866 has celestial coordinates close to WASP-12 in order to reduce differences in air mass and water vapor content in the time between the primary target and telluric observations.  In addition, the instrumental setup used two $2\arcsec$ sky fibers.  

We initially obtained 63 total exposures of WASP-12.  Of these, 12 were while the planet is transiting the star's disk.  Several (14) of these observations needed to be discarded due to low signal-to-noise ratio (S/N) or other miscellaneous reduction issues; however, only one of these 14 was an in-transit observation (discarded due to very low S/N), for a total of 38 out-of-transit exposures and 11 in-transit exposures.  An additional out-of-transit exposure was discarded in the echelle order containing \ion{Na}{1}, due to an uncertain reduction in that order.  Fig.~\ref{fig:phasecurve} indicates where the observations were taken relative to the planet's light curve as described in \citet{Hebb2009}.  Note that the out-of-transit observations are distributed in phase such that there are many observations far from transit, as well as many pre- and post-transit observations that are very close to transit.  Based on this, and the possibility of pre- or post-transit absorption, we discuss in additional detail which observations should be considered ``in" vs.~``out" of transit in \S\ref{ss:Haphase}.

\begin{figure}
\begin{center}
\epsscale{1.1}
\plotone{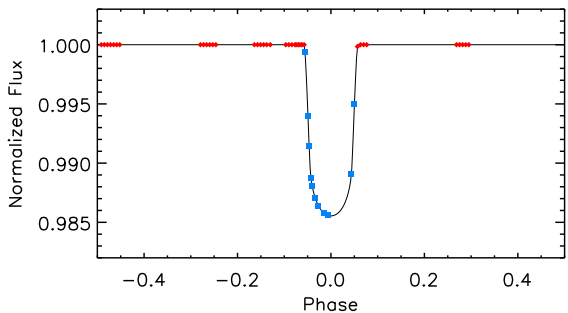}
\end{center}
\caption{This figure indicates the approximate phase placement of our WASP-12 observations, by placing them along a calculated white light transit curve using parameters from \citet{Hebb2009}.  Individual observations are classified as ``in-transit" or ``out-of-transit" based on the mid-point of the observation; out-of-transit observations are denoted by red diamonds, in-transit observations are denoted by blue squares.  Note that there are also many observations just before and after transit.}
\label{fig:phasecurve}
\end{figure}

\subsection{Data Reduction}
\label{ss:reduction}
The HRS instrument on the HET is an echelle spectrograph.  We used standard IRAF procedures for the removal of the bias level and scattered light, field flattening, aperture tracing, and wavelength calibration from the ThAr comparison lamp exposures.  We then summed the apertures to one-dimensional spectra for the primary target (WASP-12), telluric standard (HR 2866), and ThAr lamp exposures.  The ThAr lamp exposures were used to set the wavelength scale of the primary and telluric observations.

An attempt was made to extract the sky fiber apertures to one-dimensional spectra.  However, the sky background was very often faint enough that a conventional aperture trace through standard IRAF procedures failed (the trace was lost).  As a result, we did not complete the sky extraction or perform sky subtraction, and we evaluated through different methods whether or not a variable sky background impacted our final measurements.  This is discussed more in \S\ref{ss:contamination}.

After the extraction of the apertures, each relevant order of each observation was normalized using a high-order spline function to remove the blaze function of the echelle.  Flux errors are calculated based on photon and read noise, and scaled by this normalization.

\section{ANALYSIS}
\label{s:analysis}
\subsection{Transmission Spectra}
\label{ss:transmission}
The primary goal of our measurements is to determine the ``transmission spectrum" of WASP-12b:
\begin{equation}
\label{eq:transspec}
S_T=\frac{\fmeas}{\fref}-1,
\end{equation}
where $S_T$ is the transmission spectrum, $\fmeas$ is a measurement of the flux that we want to consider, and $\fref$ is an appropriate reference flux.  We will refer to either $\fmeas$ or $\fref$ generally as a ``flux spectrum" to distinguish it from the transmission spectrum.  Note that by this terminology a flat ``flux spectrum" is normalized to a value of one while a flat ``transmission spectrum" is normalized to a value of zero.  Commonly, e.g.~in Papers I--III, $\fmeas=F_{\rm in}$ (i.e., the in-transit flux) and $\fref=F_{\rm out}$ (i.e., the out-of-transit flux).  We will refer to a transmission spectrum that includes all observations categorized as either ``in" or ``out" as a ``master" transmission spectrum, though in \S\ref{ss:Haphase} we will discuss whether it is appropriate to determine this solely by the transit of the planet's disk.

In order to create a flux spectrum ($\fmeas$ or $\fref$) that will be used to calculate a transmission spectrum, we take the normalized individual spectra and coadd them (weighted by the normalized flux errors).  Prior to coaddition, simple profile fits of the stellar lines are performed in order to determine the line centroids; spectra are then shifted linearly to align the stellar features.  Higher-order resolution and wavelength corrections that were applied in Papers I--III are not necessary due to the $\sim 4\times$ lower resolution of the WASP-12b observations.  Furthermore, as explored in Paper II, observed time-based variations in the resolution of the HET HRS decreased significantly after December 2007.

The flux spectra, $\fmeas$ and $\fref$, are then used to create the transmission spectrum $S_T$ as in Equation \ref{eq:transspec}; this is done for the orders of interest containing H$\alpha$ (6563 \AA{}), H$\beta$ (4861 \AA{}), \ion{Na}{1} (5890 and 5896 \AA{}), and \ion{Ca}{1} (6122 \AA{}).  Unlike the targets in Papers II and III, the instrumental setup of the WASP-12 observations was shifted to shorter wavelengths to include H$\beta$, at the expense of potentially exploring \ion{K}{1} (7699 \AA{}).

We also define the equivalent width, $\eqw$, in terms of a transmission spectrum to be:
\begin{equation}
\label{eq:eqw}
\eqw=\int S_T \; d\lambda,
\end{equation}
where the integration is carried over some appropriate wavelength range; $\eqw$ has wavelength units.  We then use $\eqw$ as a measurement of merit for a given transmission spectrum.  Obviously, $\eqw=0$ indicates a transmission spectrum that is flat on average, while $\eqw<0$ indicates net absorption and $\eqw>0$ indicates net emission.

Our definitions of $\fmeas$ and $\fref$ are more general in Equation \ref{eq:transspec} than in the examples of Papers I--III in order to allow for variations of the transmission spectrum as a function of phase that may not necessarily occur during the disk transit of the planet.  For example, \citet{Cauley2015, Cauley2016} calculated a light curve of H$\alpha$ absorption (equivalent widths $\eqw$) as a function of phase in order to observe the possibility of pre- and/or post-transit absorption.  Each H$\alpha$ $\eqw$ is based on a corresponding ``transmission spectrum," even where $\fmeas$ is not necessarily calculated from in-transit spectra.  Again, this is in contrast with Papers I--III, where the focus was on the master transmission spectrum as defined previously.  The difference between these two sets of papers was due to the lower S/N of the HET observations compared to the Keck observations, as well as the limited periods of time for which the HET can track an object; no single night of our HET observations provides a complete transit with adequate out-of-transit baseline time.  We note, however, that single-exposure transmission spectra were considered briefly in Paper III.  In the current paper, we will examine the transmission spectrum as a function of phase and also consider the aggregate, master transmission spectrum; results are presented in \S\ref{s:results}.

\subsection{``Empirical Monte Carlo" Error Analysis}
\label{ss:EMC}
The process of normalizing individual spectra, coadding them to create flux spectra, and creating a transmission spectrum involves many possible systematic effects.  In order to quantify our transmission spectrum errors, in Papers I--III, and \citet{Cauley2015, Cauley2016}, we performed an ``Emprical Monte Carlo" (EMC) analysis of our results.  The EMC is similar to a ``jackknife" method \citep[Paper II,][]{Wall2003}.  In short, the fundamental process of the EMC is to select a subset of individual spectra, coadd them to create the $\fmeas$ flux spectrum, and then compare $\fmeas$ to an appropriate $\fref$ flux spectrum to create the transmission spectrum as defined in Equation \ref{eq:transspec}.  The transmission spectrum is then characterized by a measurement of $\eqw$ as in Equation \ref{eq:eqw}.  The distribution of $\eqw$ measuements, carried out over many different subset combinations, then is used to estimate the uncertainty in the ``master" measurement (the $\eqw$ of the $S_T$ from comparing the master in-transit spectrum vs.~the master out-of-transit spectrum).

In Papers I--III three variations on the EMC were used.  The ``out-out" method used random subsets of the out-of-transit exposures to create $\fmeas$ and used the the master ``out" flux spectrum as $\fref$.  The ``in-out" method used random subsets of the in-transit exposures to generate $\fmeas$ and  used $\fref=F_{\rm out}$ to calculate $S_T$.  Finally, the ``in-in" method used in-transit exposure subsets to generate $\fmeas$ and then created $S_T$ with $\fref=F_{\rm in}$.  The resulting ``out-out" and ``in-in" EMC distributions (measurements of $\eqw$) should be centered at zero, while the ``in-out" EMC $\eqw$ distribution should be centered at the same value as the $\eqw$ measurement of the master transmission spectrum.  The widths of the distributions can then be used to estimate the overall uncertainty in the measurement of the $\eqw$ measurement of the master transmission spectrum.

For WASP-12b we have 11 in-transit and 38 out-of-transit exposures (37 out-of-transit exposures for \ion{Na}{1}).  As compared to Papers I--III we perform the EMC analysis with two small but key differences.  First, for the various iterations we do not use the same master spectrum for $\fref$, but instead create $\fref$ from a subset of the appropriate spectra.  For example, using the ``out-out" method we select a subset of ``out" observations to create $\fmeas$ and then we use all remaining ``out" observations (the complement of those selected for the ``in") to create $\fref$.  Note that this is how the ``in-in" and ``out-out" EMC calculations were done in \citet{Cauley2015,Cauley2016}.  Secondly, in the earlier papers the number of spectra used to generate individual measurements was selected based on the overall ratio of in-transit to out-of-transit observations, e.g., if there were 25\% as many in-transit observations as out-of-transit observations, we would maintain that same ratio for the ``in-in" and ``out-out" EMC analyses.  However, especially when considering the possibility of exospheric or pre-transit absorption, we cannot take this definition of ``in" vs. ``out" for granted; therefore, in \S\ref{s:results} we put this to the test; additionally, in that section we will also discuss the sample sizes (i.e., number of spectra) and numbers of iterations for each method.

Finally, we note that an EMC analysis which is based on a subset of a larger sample should provide a somewhat conservative error estimate.  The EMC analysis is intended to characterize both systematic and statistical errors simultaneously.  If statistical errors dominate, then the actual error in the $\eqw$ of a transmission spectrum $S_T$ should scale with the number of observations involved in $S_T$ relative to the EMC analysis.  Assuming each individual spectrum has a constant uncertainty $\sigma$, then the scaling between the error in a given transmission spectrum and the error from an EMC distribution width is as follows:
\begin{equation}
\label{eq:errors}
\sigma_{S_T}=\sigma_{\rm EMC}\times\sqrt{\frac{\frac{1}{N_{{\rm meas},S_T}}+\frac{1}{N_{{\rm ref},S_T}}}{\frac{1}{N_{{\rm meas},{\rm EMC}}}+\frac{1}{N_{{\rm ref},{\rm EMC}}}}},
\end{equation}
where $\sigma_{S_T}$ is the uncertainty we are trying to find, $\sigma_{\rm EMC}$ is the error from the EMC, and the values of $N$ are the numbers of individual measured and reference spectra used in $S_T$ and the EMC.  However, this scaling may underestimate the overall error if systematic errors are significant (i.e., that there are errors that are non-normal and/or have nonzero covariance).

\subsection{Telluric Removal in Individual Spectra}
\label{ss:telluric}
Papers I--III used certain reduction and analysis techniques such as ``cleaning" the telluric standard spectra from weak stellar and interstellar lines.  These techniques are unnecessary in the present work due to WASP-12's faintness, and the subsequent lower S/N and lower resolution as compared to the targets and instrumental setup in those papers.  In fact, this extends to the subtraction of telluric absorption itself.  We used the Molecfit program \citep{Smette2015} to fit the telluric lines seen in our observations of our telluric standard target HR 2866.  We then applied the fit model to the primary WASP-12 spectra, using the model as a basis to fit the telluric lines in the WASP-12 spectra.  This fit allows for a variation in the wavelength shift and optical depth.  However, the modeled telluric lines are generally weaker than the noise of individual WASP-12 spectra, and the resulting fit parameters for the optical depth variation are not particularly physically reasonable (e.g., optical depths more than twice the original optical depth fit of the telluric spectra taken in close spatial and temporal proximity, as the weak, narrow lines ``lock on" to features in the noise).  An example of this for H$\alpha$ is shown in Fig.~\ref{fig:tellexample}.  

\begin{figure}
\begin{center}
\epsscale{1.2}
\plotone{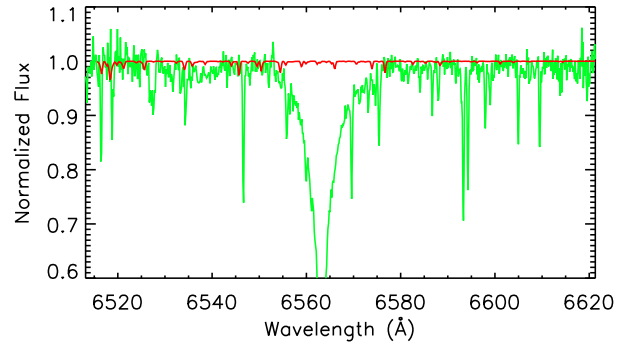}
\end{center}
\caption{Shown here is one of our WASP-12 exposures (in green) with the telluric model (in red; from a fit to a telluric observation) and the adjusted telluric model (in orange; dashed-dotted line), which is allowed to vary in depth and wavelength to fit the WASP-12 exposures.  Because the depth of most of the telluric lines are smaller than or comparable to the noise level or are otherwise blended with strong stellar features, the adjusted fit is not necessarily reasonable.  In this case, the best fit increases the optical depth by a factor of 2, the maximum it is allowed to vary.  As we argue in the text, the effect of not removing the telluric lines is ultimately minimal.}
\label{fig:tellexample}
\end{figure}


In order to test the possible effect of telluric contamination we present our first transmission spectrum, with $\fmeas=F_{\rm in}$ compared against $\fref=F_{\rm out}$ for a region with a few telluric lines to the red of the H$\alpha$ line (note that for ``in" and ``out" we use the ``pre/post" case that we will define in \S\ref{ss:Haphase}).  This is shown in Fig.~\ref{fig:WASP12telltrans}.  Note that the two strongest telluric lines in this region, at $\sim$$6571$ \AA{} and $\sim$$6575$ \AA{}, are not immediately visible in either the transmission spectrum or the reference spectrum.  A corresponding EMC analysis (Fig.~\ref{fig:WASP12tellEMC}) indicates marginal surplus flux over this limited region, approximately $1\sigma$ different than zero, where $\sigma$ is estimated from the width of the out-out EMC distribution.  The fact that a slight, statistically insignificant excess is observed rather than a deficit indicates that foregoing telluric subtraction will not result in a false positive for absorption at our lines of interest (H$\alpha$, H$\beta$, \ion{Na}{1}, and \ion{Ca}{1}) that we discuss in \S\ref{s:results}.  The width of telluric lines versus our target lines must also be considered, as the telluric lines are narrower than the stellar lines and what we might expect for the planetary atmosphere lines.

\begin{figure}
\begin{center}
\epsscale{1.2}
\plotone{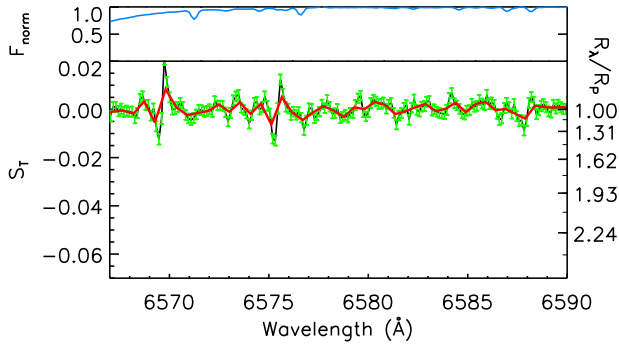}
\end{center}
\caption{The WASP-12b transmission spectrum of a region redward of H$\alpha$ that includes two telluric lines at $\sim$$6571$ \AA{} and $\sim$$6575$ \AA{}.  The figure follows the format of the corresponding figures in Paper II and III---the transmission spectrum is in gray with green errors; a binned spectrum is in red; the master out-of-transit spectrum is at the top in blue, compressed to fit on the figure.  Note that the telluric lines are not obvious in either the transmission spectrum or the reference spectrum.  Also note that the y-axis scale is chosen for consistency with similar transmission spectra figures throughout this paper.}
\label{fig:WASP12telltrans}
\end{figure}

\begin{figure}
\begin{center}
\epsscale{1.2}
\plotone{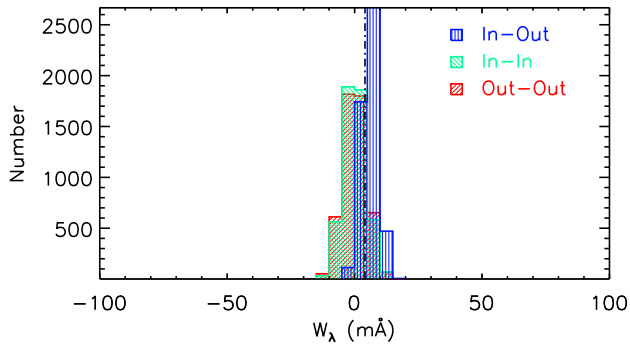}
\end{center}
\caption{Empirical Monte Carlo (``EMC") results for the telluric spectral region of in WASP-12b shown in Fig.~\ref{fig:WASP12telltrans}.  The different histograms, indicated by different colors and cross-hatching, represent the various EMC methods, described in \S\ref{ss:EMC}.  The value is the integrated transmission signal for each trial.  The dashed-dotted vertical line represents the master transmission signal, which should correspond to the centroid of the in-out histogram.  The measured absorption signal that the histograms represent is a 4 \AA{} integration that covers two strong (relative to the surrounding spectral region) telluric lines, and normalized relative to another 4 \AA{} bin to the red.  Finally, note that the x-axis scale is chosen for consistency with similar EMC figures throughout this paper.}
\label{fig:WASP12tellEMC}
\end{figure}

\subsection{Stellar and Solar Contamination}
\label{ss:contamination}
In addition to understanding the effect of the sky background (\S\ref{ss:reduction}), another consideration that must be taken into account is that WASP-12 is a hierarchical triple star system; the primary is a late F star (see discussion in \S\ref{ss:wasp12b}) orbited by a pair of M3V stars that orbit each other \citep[][and references therein]{Bechter2014}.  Our instrumental setup used the $2\arcsec$ fiber setup, which is more precisely a $1\arcsec.93$ fiber (P.~McQueen, 2017, private communication).  The separation of the dwarfs from the primary is $1\arcsec.047$, meaning that the stars' centroids will nominally be just off the edge of the fiber, and some non-trivial fraction of their flux may fall within the fiber.

In order to rigorously quantify the possible effects of inadequate sky subtraction and the stellar companions, we took modeled stellar spectra and simulated a transmission spectrum where either $\fmeas$ or $\fref$ (it does not particularly matter which) is contaminated by another stellar source.  This other stellar source is either a faint solar spectrum (to approximate the sky spectrum) or the additional M dwarf stars in the system.  These sources are scaled by an estimate of the sky background from our observations and an extrapolation of the known magnitudes of the stars, respectively.

\begin{figure}
\begin{center}
\epsscale{1.2}
\plotone{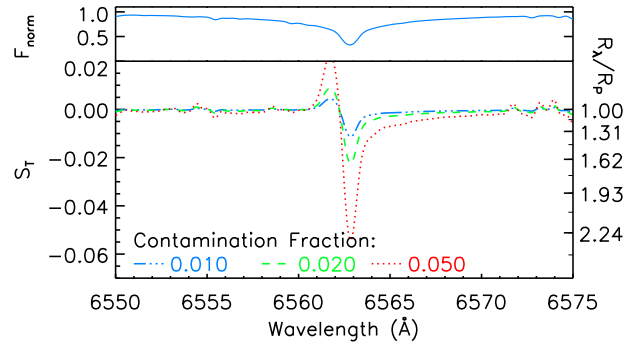}
\end{center}
\caption{A simulation of the effect of possible sky contamination, assuming that all out-of-transit observations are contaminated while in-transit observations are not.  The lower panel shows modeled contamination effects of 1\% (dash-dotted blue), 2\% (dashed green), and 5\% (dotted red).  For reference, the solid blue spectrum in the smaller upper panel shows the master stellar spectrum on a compressed scale.  We estimate our typical sky backgrounds to be 1--2\%; therefore if contamination were perfectly anti-correlated with transit (an extremely pessimistic assumption), we would see the effects on the order of the blue and green curves.  In reality, the contamination is random and the net effect is likely to be very small.}
\label{fig:sky}
\end{figure}

We can see in Fig.~\ref{fig:sky} the effect that sky contamination has on our transmission spectra.  The $18.75\kmpers$ radial velocity of WASP-12 \citep{Gaia2016} adjusted for the Earth's motion during March and April (approximately $-28\kmpers$, varying by only a few ${\rm km \; s^{-1}}$) results in a relatively constant, approximately $47\kmpers$ offset between the solar lines (causing sky contamination and to the blue) and WASP-12's stellar lines (to the red); additional RV considerations of WASP-12 due to the planetary influence are much smaller and can be safely ignored.  The simulations show the $S_T$ resulting from a 6300 K WASP-12 stellar model for both $\fmeas$ and $\fref$, but $\fref$ has been contaminated with a 5800 K solar model at 1\%, 2\%, and 5\% levels; the solar contamination has been offset according to the above discussion.  This results in an absorption feature at the wavelength of the stellar line core, but also a significant emission feature to the blue.  Importantly, this simulation assumes an extreme, unfavorable contamination level:  the sky background is $\sim1-2$\% of WASP-12's brightness in our observations, but these simulations assume a fully biased out-of-transit contamination with ``clean" in-transit observations.  For more realistic, random combinations we would reasonably expect the contamination effect to be much smaller.  We further note that in a simulation of no velocity offset between the solar and stellar lines (not shown), the similarity of the 5800 K solar model and the 6300 K WASP-12 model results in a broad, shallow feature rather than the significant narrow features shown.  While the zero-offset case is not representative of our data, it demonstrates that sky contamination for this target in our dataset cannot produce a false absorption line without the corresponding emission spike.

The same issue is explored for the companions of WASP-12 \citep{Bechter2014}.  The two companions have $J$ magnitude differences of 3.81 and 3.92 from WASP-12, respectively.  We extrapolate this to R-band flux ratios based on \citet{Ducati2001} and estimate that the flux ratio of each companion to the primary is 1.8\%.  As noted above, the stars will have centroids that are nominally just outside the edge of the fiber, reducing but not eliminating any contamination effect.  The pointing accuracy of the HET ranges from approximately $0\arcsec.2$ and $0\arcsec.5$, while the PSF FWHM ranges from about $1\arcsec.2$ to $2\arcsec.5$, depending on seeing, with a median of $1\arcsec.7$ (P.~McQueen, 2017, private communication).  These PSF FWHM ranges are consistent with the measured values in the HET night reports for our observations.  This indicates that the median flux falling in the fiber is 45\%.  The fiber will also not catch all of the flux of the primary; if perfectly centered, approximately 82\% of the primary's flux will fall in the fiber on a night of median seeing; for poor centering (and median seeing), this value drops to approximately 71\%.

In total, these values indicate that a $\sim$2\% flux ratio between the companions and the primary is a reasonable upper limit.  As in the previous simulation, the representative values in Fig.~\ref{fig:star} assume a pessimistic perfect anti-correlation of maximum contamination for the out-of-transit observations and no contamination for the in-transit signals.  No velocity offset has been assumed between WASP-12 and the companion stars; while velocity information is not available for the companion stars, all three stars in the system are resolved at a distance of approximately 430 pc \citet{Gaia2018}, implying that any orbital velocities must be small.  As before, in practice we expect a significantly smaller net contamination due to randomness.  We have also simulated the case in which contamination mimics absorption rather than emission; both are equally probable.  The lack of velocity offset results in a more straightforward absorption feature as compared to the sky contamination case; these worst-case scenarios can still be compared to our results in \S\ref{s:results}.

\begin{figure}
\begin{center}
\epsscale{1.2}
\plotone{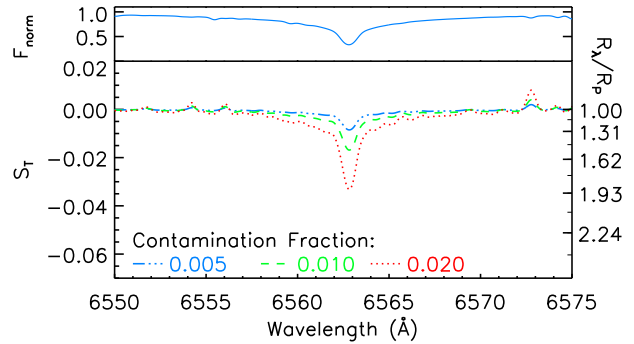}
\end{center}
\caption{Same as Fig.~\ref{fig:sky} but for possible contamination due to WASP-12's companion M dwarf stars.  The range of likely contamination fractions is discussed in the text; the 2\% curve here can be considered an extremely pessimistic upper limit on contamination.}
\label{fig:star}
\end{figure}

\subsection{Interstellar Absorption}
\label{ss:interstellar}
As noted in \S\ref{ss:telluric}, in Papers I--III it was necessary to remove interstellar absorption (as well as stellar absorption) from the telluric observations.  This is because the telluric observations are adjusted for wavelength and line depth during the process of being subtracted from the primary (using standard IRAF procedures).  However, as the stellar and interstellar lines are fixed in depth and wavelength relative to the telluric lines in the telluric spectra, making these adjustments to the telluric spectra then subtracting them from the primary spectra can introduce artifacts.

In the present work, even though WASP-12 is at a distance of approximately 430 pc \citep{Gaia2018} and interstellar lines may be significant, the lines have fixed depth and velocity offsets relative to the stellar lines of WASP-12 (RV variations for WASP-12 are small enough that they can safely be ignored).  Therefore, there is no need to remove interstellar lines prior to the calculation of $S_T$, where they will naturally be removed in the same way that stellar lines are.  In addition, while interstellar lines in general may be significant for WASP-12 due to its distance, this is not necessarily true of all the specific absorption lines explored in this paper:  interstellar lines from the \ion{Na}{1} D doublet should be present, but we would not expect detectable interstellar Balmer lines of hydrogen.

\section{RESULTS}
\label{s:results}
In this section we first present our results for a preliminary phase curve of H$\alpha$ that was calculated in order to assess whether or not there is evidence for pre- or post-transit absorption.  From this, we wish to better determine which spectra to include in $\fmeas$ and $\fref$ for our master $S_T$ and EMC analyses (\S\ref{ss:Haphase}).  Using this information, we then analyze H$\alpha$ and H$\beta$ (\S\ref{ss:halpha}), \ion{Na}{1} (\S\ref{ss:sodium}), and the \ion{Ca}{1} control line (\S\ref{ss:calcium}).  We briefly discuss the velocity shifts of observed absorption in \S\ref{ss:velocities}.

\subsection{Preliminary Phase Curve}
\label{ss:Haphase}
As noted in \S\ref{ss:wasp12b}, \citet{Fossati2010} and \citet{Haswell2012} provided evidence of metal absorption occurring in pre-transit observations of WASP-12b, with significant theoretical follow-up \citep[e.g.,][]{Lai2010,Vidotto2010,Llama2011}.  In \citet{Cauley2015,Cauley2016}, evidence for pre-transit H$\alpha$ in HD 189733b was presented.  Therefore, as discussed in \S\ref{ss:transmission}, we should not take for granted that absorption might only occur during the disk transit of WASP-12b.  To evaluate this, we create a light curve of the H$\alpha$ absorption using all out-of-transits observations to generate $\fref$ with subsets of three or more spectra, grouped by orbital phase, to generate the various $\fmeas$ and subsequent $S_T$.  Fig.~\ref{fig:HaplhaPhase} shows the $\eqw$ measurements corresponding to each $S_T$ as a function of orbital phase.

\begin{figure}
\begin{center}
\epsscale{1.2}
\plotone{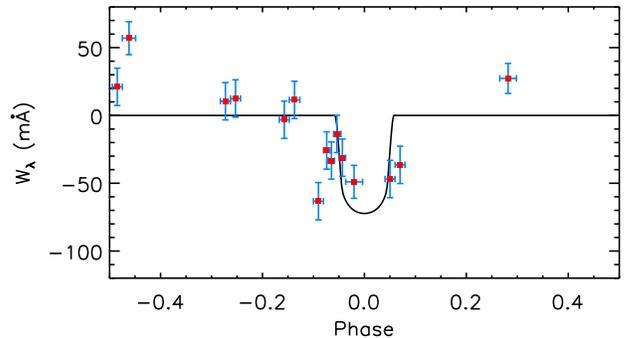}
\end{center}
\caption{A preliminary light curve of $\eqw$ measurements for H$\alpha$ absorption in WASP-12b.  The individual $S_T$ from which the $\eqw$ measurements are calculated from a $\fmeas$ of three or more individual spectra grouped by phase compared to the master out-of-transit flux spectrum as $\fref$ (the ``disk transit" case described in \S\ref{ss:Haphase}).  The $\eqw$ errors use the ``in-out" EMC widths from Table \ref{table:absorption}, scaled by the relationship in Equation \ref{eq:errors}.  The error bars on the phases respresent the range of the observations.  A calculated white light curve for WASP-12b has been added to visually indicate where the disk transit is.  In order to account for the different types of measurement (light curve vs.~$\eqw$), two adjustments have been made to the curve.  First, the baseline has been shifted from an out-of-transit value of 1 to an out-of-transit value of zero.  Second, the depth of the curve has been arbitrarily scaled to better visually match the H$\alpha$ points.}
\label{fig:HaplhaPhase}
\end{figure}

In Fig.~\ref{fig:HaplhaPhase} we see a very clear correlation with transit that extends to before and after transit.  The earliest indication of absorption is around a phase of $-0.1$, where a phase of zero is defined as the transit midpoint.  There is an indication of absorption just beyond transit at a phase of approximately 0.07.  Notably, nearly all but one of the remaining out-of-transit points (where the phase absolute value is less than 0.1) are above the baseline.  This is sensible, as it indicates that the default master out-of-transit flux spectrum used for $\fref$ has been diluted by the inclusion of the observations near transit.

In \citet{Fossati2010}, the indication of metal absorption (e.g., \ion{Mg}{2}) occurs as early as a phase of $-0.08$, which is roughly consistent with what we see here.  The early absorption is most pronounced in the NUVA band (2539--2580 \AA{}), which includes resonance lines of \ion{Na}{1}, \ion{Al}{1}, \ion{Sc}{2}, \ion{Mn}{2}, \ion{Fe}{1}, and \ion{Co}{1} \citep{Fossati2010,Morton1991,Morton2000}; a NUVA time-tag data point at a phase of approximately $-0.09$ in \citet{Fossati2010} does not show absorption.  \citet{Haswell2012} show evidence for early absorption in the NUVA and NUVC (2770--2811 \AA{}) in phases as early as $-0.16$.  As discussed in \S\ref{ss:wasp12b}, the combined results of \citet{Fossati2010} and \citet{Haswell2012} are evidence for a time-variable ingress.  It is also not a given that H$\alpha$ absorption should precisely correlate with the absorption in the NUVA and NUVC bands in those papers.

In addition to being evidence for possible pre- and/or post-transit absorption, this is an indication that we should shift our definitions of ``in-transit" and ``out-of-transit" for our calculations of $\fmeas$ and $\fref$ in either our determination of $S_T$ or when peforming our EMC analyses.  We will henceforth refer to two cases:  (1) the ``disk transit" case where we define in-transit and out-of-transit observations based on whether they occur during the white light transit of WASP-12b's disk and (2) the ``pre/post" case where we define any observation with a phase between $-0.1$ and $0.1$ to be in-transit and all others to be out-of-transit based on the combination of our preliminary light curve results and the \citet{Fossati2010} and \citet{Haswell2012} results.  Our discussion will focus primarily on the ``pre/post" case.

\subsection{H$\alpha$ and H$\beta$ Transmission Spectra}
\label{ss:halpha}
In Fig.~\ref{fig:HaplhaPhase09} we show the modified version of the light curve for H$\alpha$ assuming the pre/post case; this is the same as Fig.~\ref{fig:HaplhaPhase} but shifted, so the clear correlation with transit, including some evidence of absorption during pre- and post-transit phases, remains intact.  As discussed in \S\ref{ss:hinexo}, evidence for the circumplanetary nature of H$\alpha$ absorption in HD 189733b was presented in \citet{Cauley2017c,Cauley2017b} on the basis of short-cadence stellar H$\alpha$ monitoring and simulations of the contrast effect.  While we have not performed similar investigations of the WASP-12 system, both concepts are relevant here.  WASP-12 has a very low value of $\log{R'_{\rm HK}}$ \citep{Knutson2010}, indicating anomalously low stellar activity; however \citet{Fossati2013} argue that material in the WASP-12 system may have a significant impact on the observed core absorption in the \ion{Ca}{2} H and K lines and WASP-12 may in fact have a more normal activity level for its spectral type.  In either case, the activity level of WASP-12 is expected to be smaller than HD 189733 based on spectral type, which is relevant for conclusions about absorption in the transmission spectrum.  First, the contrast effect should be small.  Secondly, it strengthens the case that any apparent pre- or post-transit absorption, which cannot be explained by contrast effects, is due to circumplanetary absorption rather than stellar variability in the H$\alpha$ line.

\begin{figure}
\begin{center}
\epsscale{1.2}
\plotone{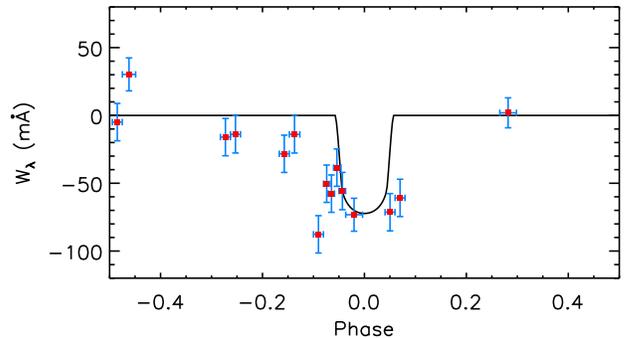}
\end{center}
\caption{The revised light curve of $\eqw$ measurements for H$\alpha$ absorption in WASP-12b.  The format is the same as Fig.~\ref{fig:HaplhaPhase}, except that this figure is for the pre/post case as defined in \S\ref{ss:Haphase}.  In addition, note that the two points above the baseline of zero are the only points that are created from more than three individual spectra, due to how the observations are binned; the point at a phase of approximately $-0.46$ is comprised of four observations and the point at a  phase of approximately $0.28$ is comprised of five observations.}
\label{fig:HaplhaPhase09}
\end{figure}

Fig.~\ref{fig:Hatrans} shows the master transmission spectrum at H$\alpha$, again assuming the pre/post case.  The presence of a line is very clear and striking, larger than the corresponding observations for other planets where H$\alpha$ has been observed (HD 189733b, KELT-9b, and KELT-20b).  The depth of the feature is stronger than the simulated worst case scenarios in either Fig.~\ref{fig:sky} or \ref{fig:star}.  In addition, no obvious, strong emission feature to the blue is seen as in Fig.~\ref{fig:sky}, though there are slight excesses to both the red and blue of the central absorption.

\begin{figure}
\begin{center}
\epsscale{1.2}
\plotone{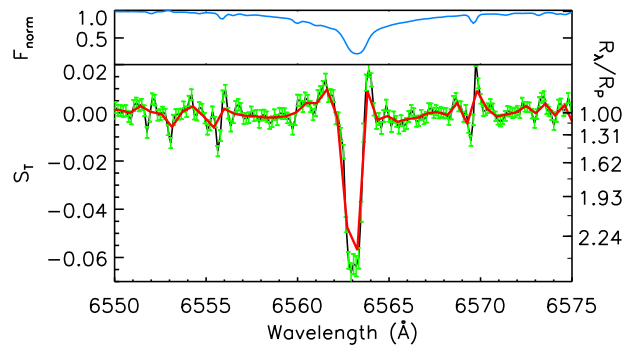}
\end{center}
\caption{The master transmission spectrum of WASP-12b at H$\alpha$.  The format is the same as Fig.~\ref{fig:WASP12telltrans}.}
\label{fig:Hatrans}
\end{figure}

The master transmission spectrum is integrated over a 2 \AA{} band, resulting in an $\eqw$ at H$\alpha$, or $W_{{\rm H}\alpha}$, of $-64.9$ m\AA{}.  We used an EMC analysis (\S\ref{ss:EMC}) to evaluate the error on the master transmission spectrum.  In the pre/post case, we have 25 in-transit and 24 out-of-transit spectra.  In our EMC analysis we select random subsets of approximately half of the available spectra for each iteration; in this case, we have 25 in-transit spectra, and thus $C^{25}_{13}=5200300$ combinations of 13 spectra are available.  This is an unwieldy number of combinations, so we choose to explore 5000 iterations for the ``in-in" method---in each iteration, $\fmeas$ is a random combination of 13 individual spectra and $\fref$ is the complement, i.e., the combination of the remaining 12 individual spectra.  Similar numbers are applicable for the ``out-out" method, with the change that there are only 24 total spectra, but we use combinations of 12 spectra to create $\fmeas$ and the complementary 12 spectra to create $\fref$; this is done for 5000 iterations (combinations).  For the ``in-out" method we choose random subsets of 13 of the in-transit spectra to create $\fmeas$ and random subsets 12 out-of-transit spectra to create $\fref$, again for 5000 iterations.

\begin{figure}
\begin{center}
\epsscale{1.2}
\plotone{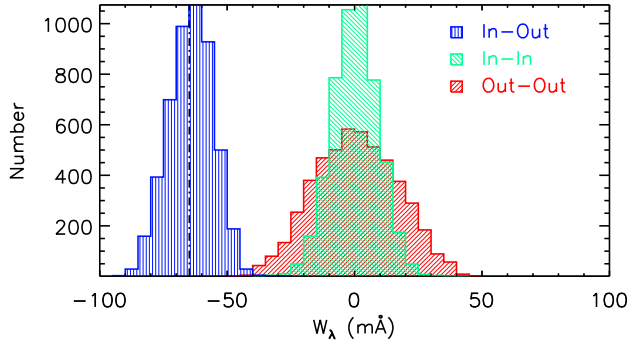}
\end{center}
\caption{Empirical Monte Carlo (``EMC") results for H$\alpha$ in WASP-12b.  The format is the same as Fig.~\ref{fig:WASP12tellEMC}.}
\label{fig:HaEMC}
\end{figure}

We show our EMC analysis in Fig.~\ref{fig:HaEMC}.  In that figure, the ``in-in" and ``out-out" distributions are both centered at zero, which is what we expect.  In addition, the ``in-out" distribution is centered at roughly the master transmission spectrum, also what we expect.

We assume that the distribution of $\eqw$ measurements is Gaussian, and allow the measured Gaussian $\sigma$ to represent an estimate of the error in our master $\eqw$ measurements, keeping in mind that we may wish to scale the errors by Equation \ref{eq:errors}.  The results of the integration of the master transmission spectrum and the errors derived from the EMC are shown in Table \ref{table:absorption}; results are given for H$\alpha$ along with H$\beta$, \ion{Na}{1}, and \ion{Ca}{1}.  For H$\alpha$ the ``out-out" distribution is broader than the ``in-in" and ``in-out."  As seen in the revised phase curve (Fig.~\ref{fig:HaplhaPhase09}) there is some variation in the baseline, which contributes to the broader width of the ``out-out" distribution.  The source of this variation is not immediately obvious, but the EMC analysis is intended to characterize systematic effects such as stellar variation or variability in the normalization of individual spectra.  We note here that in Paper II and Paper III it can be seen that the EMC distribution of strong lines such as H$\alpha$ and \ion{Na}{1} are somewhat broader than the EMC distributions of weaker lines like \ion{Ca}{1}, likely because these two factors (stellar variation and difficulty in normalization) are more significant for lines that are broader and stronger in the stellar spectrum.

\begin{deluxetable*}{lcccccc}
\tablecolumns{7}
\tabletypesize{\small}
\tablecaption{WASP-12b Absorption Results\label{table:absorption}}
\tablehead{\colhead{Line} & \colhead{Bin} & \colhead{$\eqw$} & \colhead{``In-In" $\sigma$} & \colhead{``In-Out" $\sigma$} & \colhead{``Out-Out" $\sigma$} & \colhead{Scale Factor\tablenotemark{b}} \\ \colhead{} & \colhead{(\AA{})} & \colhead{(m\AA{})} & \colhead{(m\AA{})} & \colhead{(m\AA{})} & \colhead{(m\AA{})} & \colhead{}}
\startdata
H$\alpha$ & 2.0 & $-64.9$ & $9.0$ & $9.1$ & $17.3$ & $0.714$ \\
H$\beta$ & 2.0 & $-3.1$ & $2.8$ & $2.7$ & $4.0$ & $0.714$ \\
\ion{Na}{1}\tablenotemark{b} & 2.0 & $-52.6$ & $15.2$ & $14.9$ & $29.2$ & $0.707$ \\ 
\ion{Ca}{1} & 2.0 &  $-2.3$ & $3.3$ & $2.5$ & $3.3$ & $0.714$ \\ 
\enddata
\tablenotetext{\rm a}{The factor by which the error in the master $\eqw$ measurements should be scaled based on Equation \ref{eq:errors}; e.g., the out-out width for H$\alpha$ would correspond to an actual error of $(17.3 {\rm \; m\AA{}})\times0.714=12.3$ m\AA{}.}
\tablenotetext{\rm b}{The $\eqw$ for \ion{Na}{1} is the total of both lines in the doublet, each with an integration width of 2.0 \AA{} as described in the text.}
\end{deluxetable*}

Next we present similar figures for the H$\beta$ phase curve (Fig.~\ref{fig:HbetaPhase}), master transmission spectrum (Fig.~\ref{fig:Hbtrans}), and EMC analysis (Fig.~\ref{fig:HbEMC}), all based on the pre/post case in order to be consistent with H$\alpha$.  There is no obvious correlation in the H$\beta$ phase curve with transit, and the transmission spectrum of H$\beta$ does not show any obvious absorption near line center.  There are some additional artifacts in the transmission spectrum at other wavelengths, but based on the EMC analysis, absorption at H$\beta$ is roughly consistent with zero at the $1\sigma$ level.

\begin{figure}
\begin{center}
\epsscale{1.2}
\plotone{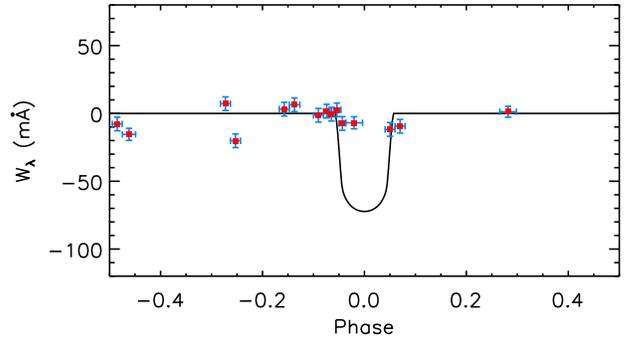}
\end{center}
\caption{Same as Fig.~\ref{fig:HaplhaPhase09}, including the in-transit and out-of-transit defintions, but for H$\beta$.}
\label{fig:HbetaPhase}
\end{figure}

\begin{figure}
\begin{center}
\epsscale{1.2}
\plotone{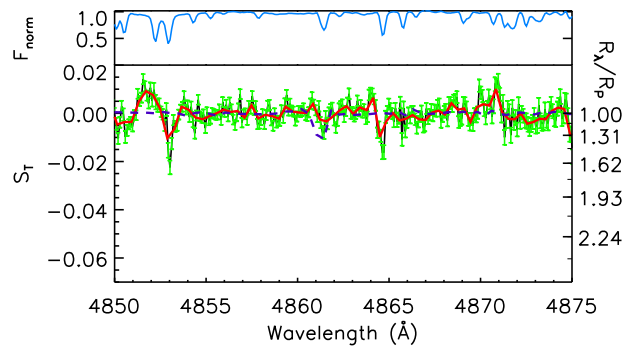}
\end{center}
\caption{The master transmission spectrum of WASP-12b at H$\beta$.  The format is the same as Fig.~\ref{fig:WASP12telltrans}.  In addition, the dotted purple line on the H$\beta$ transmission spectrum represents the binned transmision spectrum of H$\alpha$ from Fig.~\ref{fig:Hatrans}, scaled by wavelength and $f$-value to represent the expected H$\beta$ line, if optically thin.}
\label{fig:Hbtrans}
\end{figure}

\begin{figure}
\begin{center}
\epsscale{1.2}
\plotone{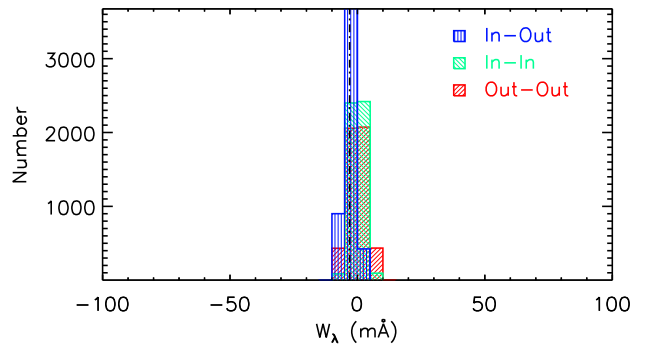}
\end{center}
\caption{EMC results for H$\beta$ in WASP-12b.  The format is the same as Fig.~\ref{fig:WASP12tellEMC}.}
\label{fig:HbEMC}
\end{figure}

Notably, \citet{Cauley2015, Cauley2016} did see H$\beta$ and H$\gamma$ absorption in Keck observations of HD 189733b, while in Paper III the HET observations did not cover these wavelengths.  However, the absorption ratios in the different Balmer lines (H$\alpha$, H$\beta$, and H$\gamma$) vary significantly between \citet{Cauley2015} and \citet{Cauley2016}.  In \citet{Cauley2015} the H$\alpha$/H$\beta$ absorption ratio is smaller, indicating some degree of optical thickness.  However, in \citet{Cauley2016} H$\beta$ is much weaker, closer to the optically thin limit, and H$\gamma$ is not observed at a statistically significant level.  The oscillator strength of the H$\beta$ transition is more than five times smaller than H$\alpha$ \citep{Goldwire1968}.  In Fig.~\ref{fig:Hbtrans} we also show the binned H$\alpha$ transmission spectrum from Fig.~\ref{fig:Hatrans}, scaled by wavelength and $f$-value.  This would represent an estimate of the expected H$\beta$ profile in the optically thin limit.  While we do not see a clear feature this large at H$\beta$, the estimate is not substantially above the remaining noise and other artifacts.

The H$\beta$ $\eqw$ measurement and EMC error estimates in Table \ref{table:absorption} suggest a very small, statistically insignificant ($\sim1\sigma$) amount of absorption at H$\beta$, which would be inconsistent with the H$\alpha$ absorption even accounting for the differing oscillator strengths (note that, in addition to $f$-value, $\eqw$ is also proportional to $\lambda^2$).  However, given the errors on both H$\alpha$ and H$\beta$, the results are not mutually exclusive at a level of approximately 2$\sigma$ if we assume the optically thin case.

\citet{Sing2013} presented a transmission spectrum of WASP-12b using {\it HST} STIS G430L and G750L.  In that paper, the transmission spectrum at H$\alpha$ is not explicitly shown.  \citet{Sing2013} instead show broadband transmission spectral results in wavebands hundreds of angstroms in width.  The bin covering H$\alpha$, from 6300--6800 \AA{}, has slightly larger transit depth than the surrounding wavebands, though not by a statistically significantly amount.  \citet{Sing2013} also searched for H$\alpha$, H$\beta$, \ion{Na}{1}, and \ion{K}{1} in narrower bandpasses and did not find any evidence for absorption.  They note that their results for \ion{Na}{1} do not rule out absorption confined to the narrow core, as observed for HD 189733b in the varying results of Papers I--II, \citet{Huitson2012}, and \citet{Sing2012}.  Our observations here resolve the core of the stellar H$\alpha$ line, which is critical for sensitivity to detecting absorption in the core.  STIS G750L, at a resolution of $R\sim500$, does not resolve the core; Fig.~\ref{fig:Hatrans} covers less than two G750L instrumental resolution elements.  It is ultimately unsurprising that the H$\alpha$ signal we see here is not observed in the {\it HST} data.

In \S\ref{ss:planetology} we discussed the differences between the WASP-12 and HD 189733 systems.  At the time of its discovery, WASP-12b was the most highly irradiated exoplanet known.  However, as a late F star WASP-12 is expected to be less active and have less Ly$\alpha$ emission than the K0V star HD 189733.  To explain the H$\alpha$ absorption in HD 189733b, \citet{Huang2017} modeled the $2\ell$ population in the atomic hydrogen layer between pressures of $5\times10^{-5}$ $\mu$bar and 10 $\mu$bar, high in the atmosphere \citep[for more discussion see][]{Yelle2004}.  This model indicates that radiative excitation from Ly$\alpha$ dominates over excitation mechanisms for hydrogen.  Thus, it is superficially surprising to see H$\alpha$ absorption in WASP-12b if stellar Ly$\alpha$ is the dominant driver of the $n=2$ hydrogen population.

One possibility is that WASP-12b has insufficient cooling, perhaps due to an underabundance of coolants, which would allow for an increased $n=2$ hydrogen population even with limited Ly$\alpha$ flux coming from the star.  However, the \citet{Fossati2010} and \citet{Haswell2012} results indicate that magnesium, an important coolant, is present in the upper atmosphere.  To model this issue in detail is beyond the scope of this paper, but our results provide motivation for undertaking such modeling.  We note here that the HD 189733b models of \citet{Huang2017} and \citet{Christie2013} find that H$\alpha$ absorption occurs over a relatively small range of radii.  However, our observations---both the magnitude of the absorption and the evidence for pre- or post-transit absorption---indicate that H$\alpha$ absorption occurs over a larger range of radii.  This difference may be important for understanding the mechanism of $n=2$ creation.  While we do not have a detailed model for understanding the H$\alpha$ absorption, the broader strokes of our observations, especially the early ingress times and the implication that absorption occurs at very high altitudes, are not surprising in light of the previous observational results \citep{Fossati2010,Haswell2012,Sing2013}.

The lack of observed H$\beta$ absorption is the biggest challenge to our interpretation of the H$\alpha$ transmission spectrum as showing absorption as a real, astrophysical signal from a circumplanetary source.  However, there is some possibility that we have underestimated errors for the H$\beta$ non-detection in particular, and even with our current errors the two results are not inconsistent at high statistical significance.  What we can conclude is that the lack of clear H$\beta$ absorption suggests that the H$\alpha$ absorption in WASP-12b, if it is of circumplanetary origin rather than some sort of artifact, is optically thin.  Ultimately, it is not clear if this represents a difference with respect to HD 189733b or not.

\subsection{\ion{Na}{1} Transmission Spectrum}
\label{ss:sodium}
Our results for \ion{Na}{1} in the pre/post case are shown in Figs.~\ref{fig:NIPhase} (the phase curve), \ref{fig:Natrans} (master transmission spectrum), and \ref{fig:NaEMC} (EMC analysis).  Our master transmission spectrum $\eqw$ is $-52.6$ m\AA{}, and significant to $\sim5\sigma$ using either the in-in or in-out EMC distribution widths for the error (and scaled by a factor of 0.707), but approximately $2.5\sigma$ using the very broad, scaled out-out width.  As with H$\alpha$, it is not immediately obvious why the out-out width is so broad, other than we see in the phase curve that there is significant variation in the out-of-transit observation.  In the disk transit case (not shown), absorption is still observed albeit with a weaker value, and an EMC that is similarly broad, such that the overall measurement has marginal statistical significance ($\sim1\sigma$) at best.

\begin{figure}
\begin{center}
\epsscale{1.2}
\plotone{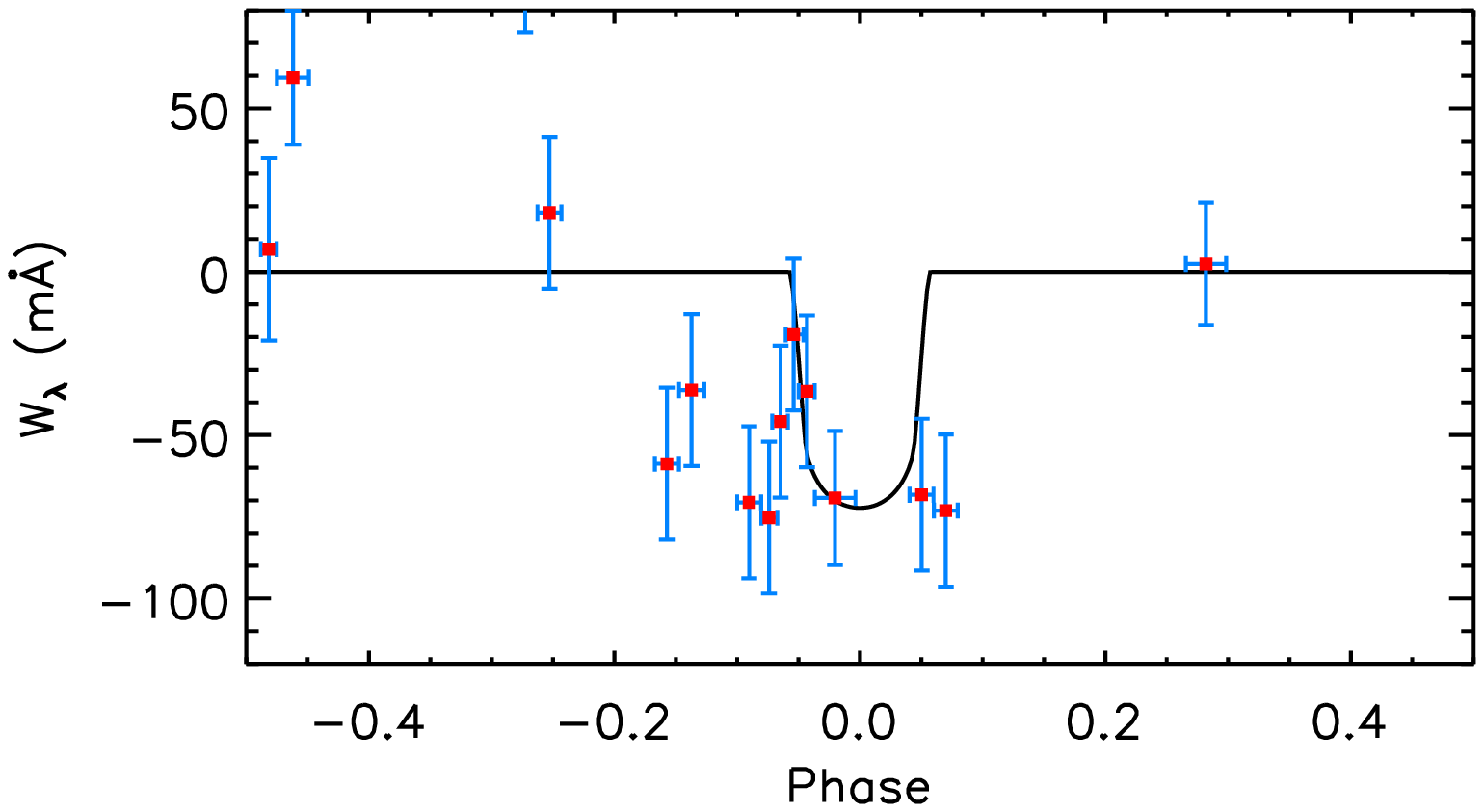}
\end{center}
\caption{Same as Fig.~\ref{fig:HaplhaPhase09} but for \ion{Na}{1}.}
\label{fig:NIPhase}
\end{figure}

\begin{figure}
\begin{center}
\epsscale{1.2}
\plotone{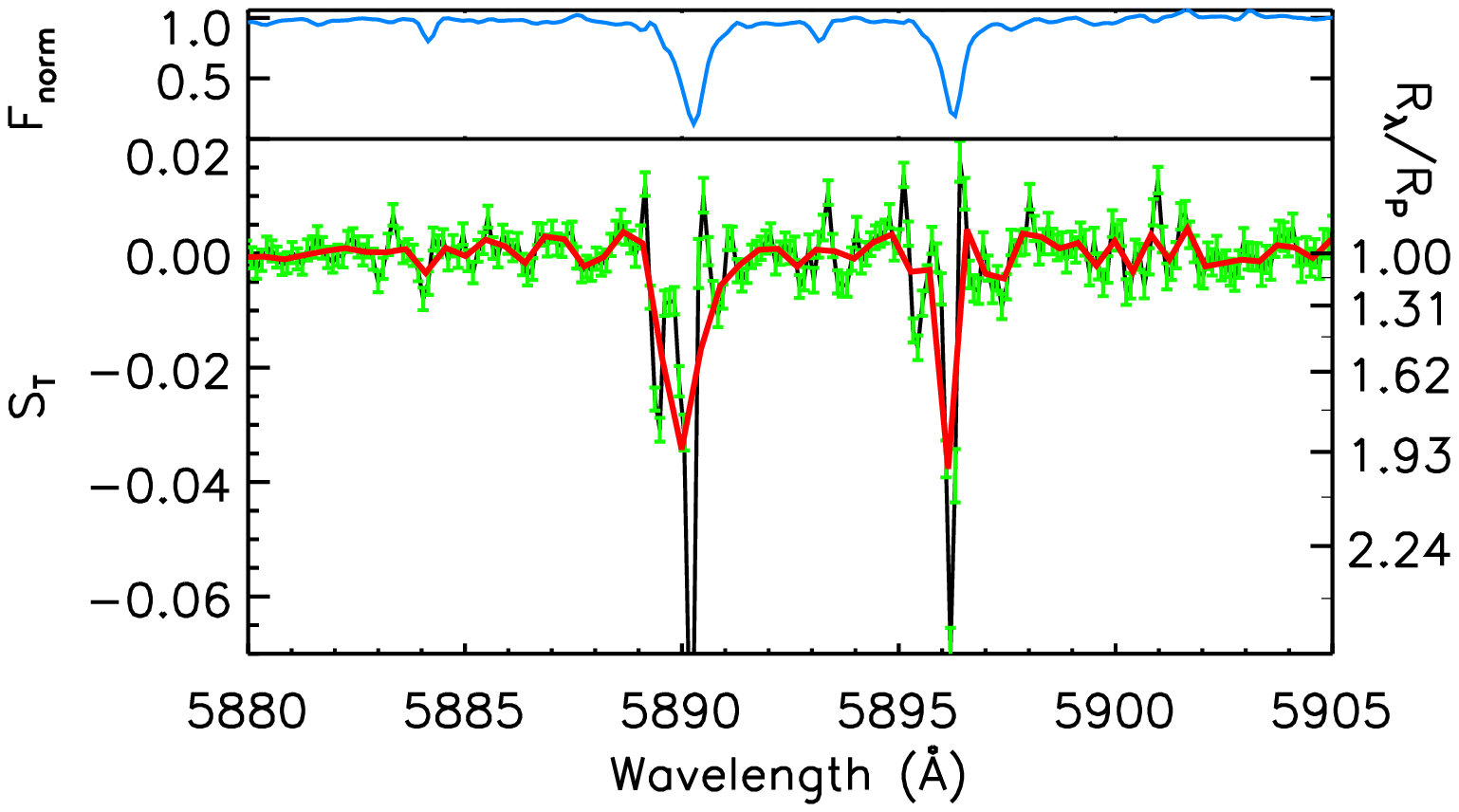}
\end{center}
\caption{The master transmission spectrum of WASP-12b at the \ion{Na}{1} D doublet.  The format is the same as Fig.~\ref{fig:WASP12telltrans}.}
\label{fig:Natrans}
\end{figure}

\begin{figure}
\begin{center}
\epsscale{1.2}
\plotone{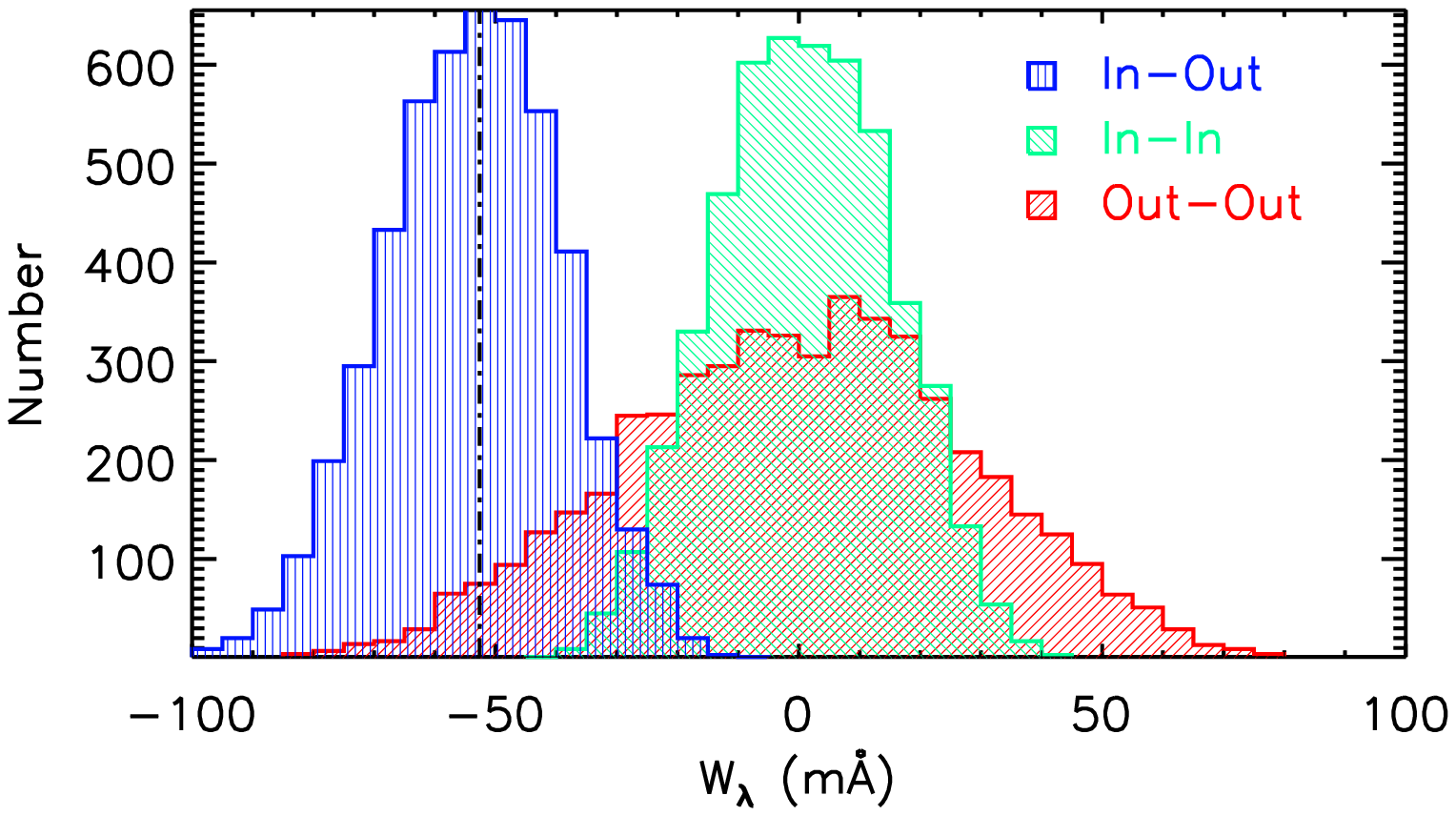}
\end{center}
\caption{EMC results for \ion{Na}{1} in WASP-12b.  The format is the same as Fig.~\ref{fig:WASP12tellEMC}.}
\label{fig:NaEMC}
\end{figure}

\citet{Burton2015} used defocused transmission spectroscopy to make a tentative detection of \ion{Na}{1} in WASP-12b, using a 2 \AA{} integration window over each line of the doublet, initially finding an absorption fraction of $0.12\pm0.03$\%.  This value is revised to $0.15\pm0.05$\% after an attempt to remove a systematic feature in mid-transit; the final reported value is $0.12\pm0.03[+0.03]$\%, with the additional error in brackets reflecting the potential effect of this systematic issue.

Converting the absorption value in Table \ref{table:absorption} from equivalent width to percentage absorption, we find a value of $0.59$\% (total for both lines of the doublet) using the disk transit criteria and 1.32\% if we include pre- and post-transit up to phases of $\pm0.1$.  However, these values are not directly comparable to the \citet{Burton2015} result.  Our method takes an average of the transmission spectrum as follows:
\begin{equation}
\label{eq:avgts}
\langle S_T \rangle_{\rm (this \: paper)} =\bigg \langle \frac{F_{in}}{F_{out}}\bigg \rangle -1 
\end{equation}
In contrast, for a single in-transit or out-of-transit spectrum, \citeauthor{Burton2015} integrate and compare their integration window to surrounding continuum windows, and then compare the in-transit and out-of-transit integrations.  In essence, their average transmission spectrum can be described as:
\begin{equation}
\label{eq:avgtsB15}
\langle S_T \rangle_{\rm (Burton \: et \: al.)} =\frac{\langle F_{in} \rangle}{\langle F_{out} \rangle}-1
\end{equation}
The difference in the order of operations of the averages is not trivial.  The \citet{Burton2015} method will typically result in smaller values because our point-by-point values of $S_T$ will have a larger absolute value in the line cores where $\fref$ is small.  When performing the integration in the manner of \citet{Burton2015}, we get an absorption value of 0.18\% across the two lines of \ion{Na}{1} in the disk transit case, reasonably consistent with their value---although, as mentioned, our disk transit result is of marginal statistical significance.  It is more appropriate to use our disk transit case results for this comparison because \citet{Burton2015} do not indicate that they include any pre- or post-transit values in their calculation.  However, our pre/post case result, integrated in this same manner, gives us an absorption value of 0.56\%.  Errors on this value will scale with the errors in the EMC results in Table \ref{table:absorption}.

\subsection{\ion{Ca}{1} Transmission Spectrum}
\label{ss:calcium}

We use the \ion{Ca}{1} line at 6122 \AA{} as a control line where planetary absorption is not expected, as was done in Papers I--III.  This is based on the assumption that \ion{Ca}{1} will condense out of the planets' atmospheres based on the pressure--temperature relationships for brown dwarfs given by \citet{Lodders}.  WASP-12b has a higher temperature than any of the targets in those papers, so first we must evaluate this assumption.  \citet{Stevenson2014} derive a terminator temperature of $1870\pm130 \rm{\; K}$.  Based on \citet{Lodders}, we estimate that \ion{Ca}{1} will not condense to CaTiO$_3$ at this temperature as long as the pressure is greater than $\sim$$0.1$ bar.  \citet{Stevenson2014} do not explicitly specify the reference pressure for this temperature, but note elsewhere that they usually set the reference pressure at $\sim$$1$ bar.  We also note that \citet{Stevenson2014} do find alternate model retrievals that fit their data with similar $\chi^2$ for higher temperatures and relatively lower reference pressures, but they ultimately favor the lower temperature.

In any case, we do not see any clear evidence for transit-correlated absorption in either the phase curve (Fig.~\ref{fig:CaIPhase}) or the transmission spectrum for \ion{Ca}{1} (Fig.~\ref{fig:Catrans}), as the latter shows no obvious features.  There is a small dip near line center that is largely washed out by 4-pixel binning, and the integration is consistent with zero at the $\sim1\sigma$ level in our EMC error analysis (Fig.~\ref{fig:CaEMC}).  These figures show the pre/post case, but the disk transit case (not shown) is not substantially different; any apparent absorption is even weaker and statistically insignificant.

\begin{figure}
\begin{center}
\epsscale{1.2}
\plotone{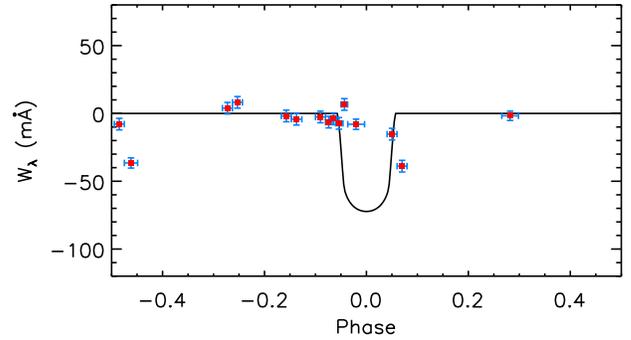}
\end{center}
\caption{Same as Fig.~\ref{fig:HaplhaPhase09} but for the \ion{Ca}{1} control line.}
\label{fig:CaIPhase}
\end{figure}

\begin{figure}
\begin{center}
\epsscale{1.2}
\plotone{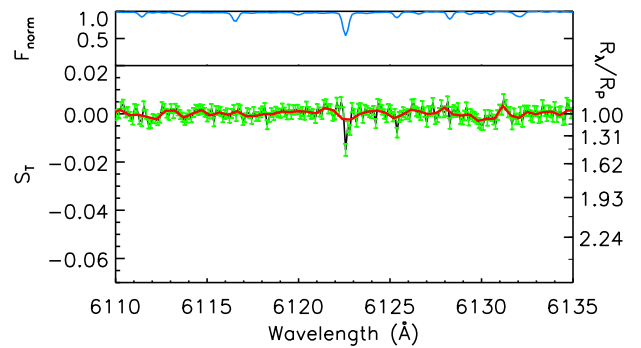}
\end{center}
\caption{The master transmission spectrum of WASP-12b at \ion{Ca}{1}, our control line.  The format is the same as Fig.~\ref{fig:WASP12telltrans}}
\label{fig:Catrans}
\end{figure}

\begin{figure}
\begin{center}
\epsscale{1.2}
\plotone{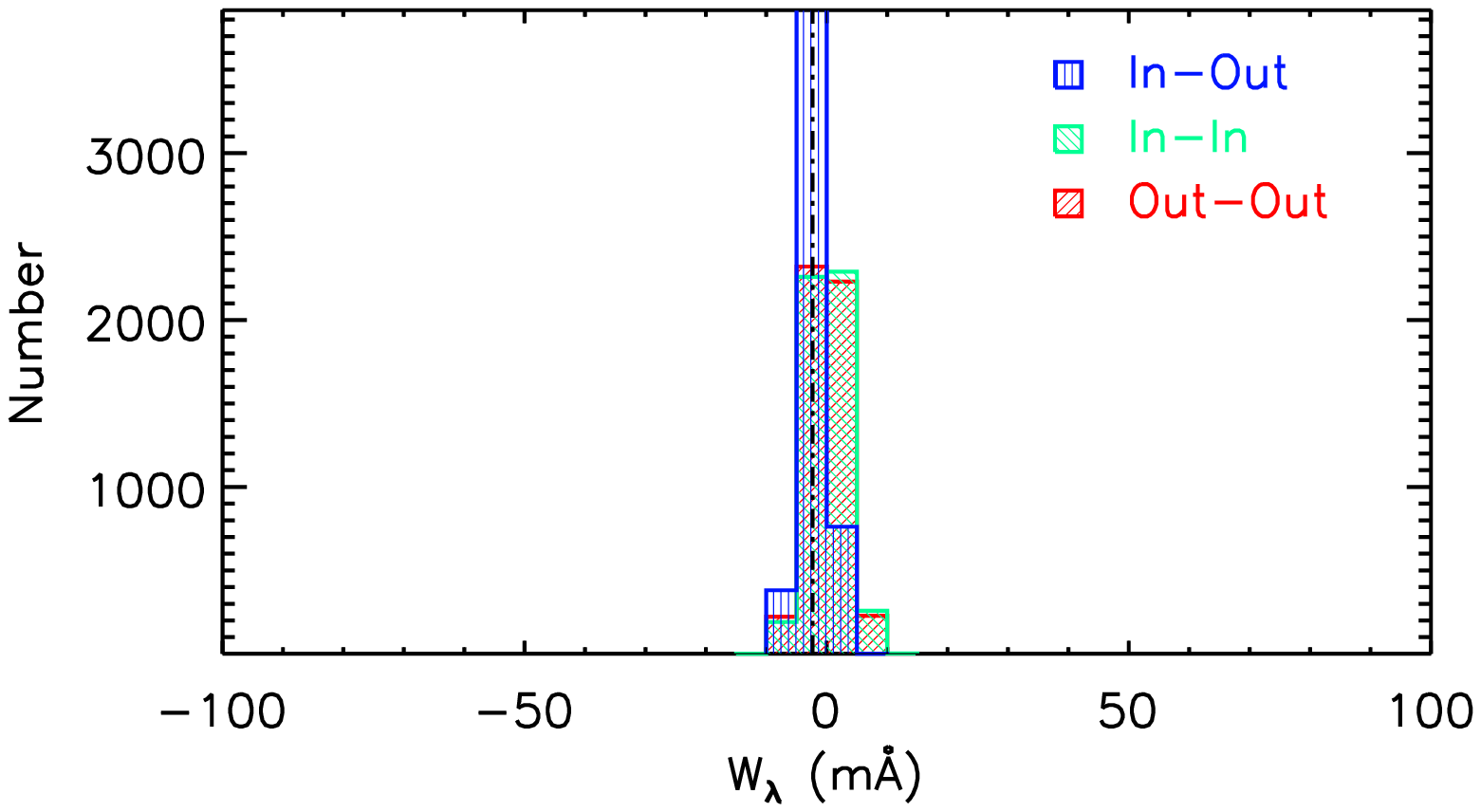}
\end{center}
\caption{EMC results for \ion{Ca}{1} in WASP-12b.  The format is the same as Fig.~\ref{fig:WASP12tellEMC}.}
\label{fig:CaEMC}
\end{figure}

\subsection{Absorption Velocities}
\label{ss:velocities}
As described in \S\ref{ss:obsdescription}, we used the HET's HRS at its $R\sim15000$ setting.  This corresponds to a velocity resolution of $\sim20\kmpers$ near H$\alpha$ and \ion{Na}{1}.  As shown in Fig.~\ref{fig:phasecurve}, the observations are weighted toward the first half of transit.  We calculate that for a simple filled Roche lobe model, assuming that hydrogen absorption matches the planet's radial velocity, any absorption should be dominated by observations corresponding to approximately $-50\kmpers$ (relative to the star).  In Figs.~\ref{fig:Hatrans} and \ref{fig:Natrans}, the H$\alpha$ and \ion{Na}{1} absorption is slightly blueshifted relative to the stellar lines, by $-9\kmpers$ (this is the value for the H$\alpha$ line and the average value for the two \ion{Na}{1} lines).  The widths are on the order of the instrumental resolution, which is smaller than the full range of velocities corresponding to the observations across the transit.  \citet{Salz2016} also note a very broad velocity range for their simulated Ly$\alpha$ transmission spectrum profile.

There are some caveats worth noting here.  The first is that our analysis method is geared toward maximizing S/N in the final $S_T$ rather than preserving velocity information.  In the coaddition step, individual spectra are self-aligned based on the stellar features (\S\ref{ss:reduction}).  It is therefore possible for this step to lose velocity information about the planet's atmosphere; that is, adding the profile of planetary absorption to the stellar line will influence the centroids of the stellar lines that we align.  In addition, upper atmospheres are dynamic and there may be other contributions to the line profile other than the planet's orbital velocity \citep[e.g., see][]{RauscherMenou2013}.  There is significant evidence that the Roche lobe of the planet is overfilled \citep{Lai2010, Debrecht2018}; while this overfilled Roche lobe is certain to include hydrogen, it is not clear that significant $n=2$ hydrogen is present throughout this envelope.

\section{Conclusions}
\label{s:conclusions}

\subsection{Discussion}
\label{ss:discussion}
As we have stated, detection of H$\alpha$ in exoplanetary atmospheres is rare.  The H$\alpha$ in HD 189733b \citep[Paper III,][]{Cauley2015,Cauley2016} is by far the best studied.  There are also a handful of non-detections of H$\alpha$ such as for HD 209458b in \citet{Winn2004} and HD 147506b and HD 149026b in Paper III.  Recent additional detections in KELT-20b \citep{Casasayas2018} and KELT-9b \citep{Yan2018} are intriguing, especially considering that they are around A stars, but they are not as well-studied as HD 189733b.

\citet{Salz2016} modeled the atmosphere of several hot Jupiter planets and found a mass-loss rate for WASP-12b that is large (3.4\% per Gyr) but significantly smaller than previous results \citep{Li2010,Lai2010}, thus resolving the prior implication that the planet was short-lived and its discovery statistically improbable.  The mass-loss rate found by \citet{Salz2016} indicates a very large hydrogen envelope, and a large theoretical Ly$\alpha$ signal that absorbs over 80\% of the incoming flux at line center and spans hundreds of $\kmpers$.  This model includes absorption that is significant well beyond the Roche lobe.  Given the abundance of hydrogen, the detection of H$\alpha$ is not particularly surprising, nor is the fact that we detect pre- and possibly post-transit absorption.  The primary open question remains what mechanisms generate $n=2$ in significant amounts, and whether these mechanisms can explain the observational details, particularly the depth and velocity range, of the H$\alpha$ absorption.

What creates $n=2$ hydrogen in hot Jupiter atmospheres?  The HD 189733b H$\alpha$ detection has been modeled in detail by \citet{Christie2013} and \citet{Huang2017}.  The former used a hydrostatic atmosphere model, but the Ly$\alpha$ radiation was not modeled in detail.  \citet{Christie2013} found that a reasonably constant $n=2$ hydrogen density within the atomic layer could be created if collisional excitation dominated.  \citet{Huang2017} expanded on the work of \citet{Christie2013} by including a more detailed treatment of the Ly$\alpha$ radiative transfer.  By considering Ly$\alpha$ coming from recombinations within the atmosphere, the radiative excitation rate can exceed the collisional excitation rate, meaning that the excitation can occur at significant levels deeper within the atmosphere at greater $n=1$ hydrogen densities.

Given that WASP-12's Ly$\alpha$ emission is presumably weaker than HD 189733's, it is unclear how significant $n=2$ could exist.  However, WASP-12b's atmosphere should be more extended than HD 189733b's, with a larger scale height.  WASP-12b's lower atmosphere is certainly hotter than HD 189733b's, though this may not necessarily be the case at higher altitudes \citep{Salz2016}.  If and where WASP-12b's atmosphere is hotter, it is possible that even with weaker EUV radiation, collisions will contribute significantly to $1s\rightarrow2s$ excitation.

Our results also must be understood in the context of the evidence for a torus- or disk-like structure around WASP-12, made up of material from WASP-12b \citep{Lai2010,Haswell2012,Fossati2013,Debrecht2018}.  We show a clear correlation with transit for H$\alpha$ and \ion{Na}{1}, which argues for some asymmetry in the disk, such as a collision between the accretion stream from the planet and the disk \citep{Lai2010}, which could be the source of early ingress.  Understanding the early ingress also has significant implications as a potential probe of the magnetic field \citep{Vidotto2010,Llama2011}.  The evidence for late egress in our observations is not as compelling as the evidence for early ingress, but it is also worth further study.

In \S\ref{ss:hinexo} we noted that there are observational challenges to relying on Ly$\alpha$ as a diagnostic of extended exoplanetary atmospheres.  First, UV instrumentation that covers Ly$\alpha$ at adequate S/N and resolution is essentially restricted to {\it HST} at the current time.  Second, the stellar flux at Ly$\alpha$ varies significantly as a function of spectral type.  Active, late-type stars have significant line emission at Ly$\alpha$; this includes K stars like HD 189733b.  Late F stars like WASP-12 have line emission at Ly$\alpha$ but it is weaker (relative to the star's continuum) than Ly$\alpha$ emission for K or M stars, while earlier type stars like the A stars KELT-9b and KELT-20b will have significant UV continuum flux.  Third, even with adequate UV instrumentation and stellar flux at Ly$\alpha$, the observation of Ly$\alpha$ transit absorption is complicated by interstellar absorption and airglow in the Earth's atmosphere, issues that are not present for H$\alpha$.

Given that H$\alpha$ also probes the structure of a planet's upper atmosphere and its location in the visible red portion of the spectrum makes it accessible from the ground, H$\alpha$ is an arguably underutilized diagnostic for observing extended atmospheres.  Specifically, while not all exoplanets with observable Ly$\alpha$ signatures will have corresponding, observable H$\alpha$ absorption (e.g., HD 209458b), using both can be complementary approaches insofar as Ly$\alpha$ observations are flux limited.  This is the case with WASP-12b; at a distance of approximately 430 pc \citep{Gaia2018}, interstellar absorption prevents Ly$\alpha$ observations of WASP-12 with {\it HST}.  The recent \ion{He}{1} detection by \citet{Spake2018} is another way in which extended atmospheres may potentially be observed without UV observations of Ly$\alpha$.  Furthermore, the far- and near-UV metals line observations discussed in \S\ref{ss:wasp12b} represent another significant way in which extended atmospheres may be probed.  In addition to being in different wavebands, all of these methods represent at least somewhat different diagnostics of exoplanetary atmospheres, and therefore they are all useful as different probes of exoplanetary characteristics.

\subsection{Summary}
\label{ss:summary}
We have presented the transmission spectrum of WASP-12b from HET observations in March/April 2012.  The spectrum shows clear features at H$\alpha$ (Fig.~\ref{fig:Hatrans}) and \ion{Na}{1} (Fig.~\ref{fig:Natrans}) while no obvious features are observed at \ion{Ca}{1} (Fig.~\ref{fig:Catrans}, intended as a control line) or H$\beta$ (Fig.~\ref{fig:Hbtrans}).  The H$\alpha$ absorption marks only the fourth such detection in an exoplanetary atmosphere after HD 189733b, KELT-20b \citet{Casasayas2018}, and KELT-9b \citep{Yan2018}, while the \ion{Na}{1} absorption is roughly consistent with previous results by \citet{Burton2015} and likewise one of larger, but still limited, number of exoplanetary \ion{Na}{1} detections.  Phase curves of H$\alpha$ (Fig.~\ref{fig:HaplhaPhase09}) and \ion{Na}{1} (Fig.~\ref{fig:NIPhase}) indicate the possibility of pre- and post-transit absorption, roughly consistent with the results of \citet{Fossati2010} and \citet{Haswell2012} for metals in the extended atmosphere of WASP-12b.  The evidence for pre- and post-transit absorption suggests that stellar effects such as the contrast effect are not an adequate explanation for our observations.  The lack of H$\beta$ absorption at a level that would be consistent with the strong H$\alpha$ signal is somewhat puzzling and the biggest challenge to our interpretation of the H$\alpha$ transmission spectrum as a real, astrophysical signal from WASP-12b's circumplanetary material; however, our results for the H$\alpha$/H$\beta$ ratio are not ruled out to high statistical significance in the optically thin case.

\subsection{Future Work}
\label{ss:future}
As with HD 189733b, the transit-correlated H$\alpha$ absorption cannot be fully understood with our track-limited observations made by the HET.  The H$\alpha$ profile of WASP-12b needs to be observed at high S/N over an entire single transit in order to get a better picture of what the correlation between transit and absorption actually is.  Our previous work has shown that absorption detected through the HET can be well-characterized through a large telescope that observes continuously through transit, e.g., with Keck \citep{Cauley2015, Cauley2016}.  Such observations would also allow us to more rigorously characterize the suggestion of pre- and post-transit absorption in our observations, as was noted in HD 189733b by \citet{Cauley2015,Cauley2016}.

\acknowledgements
This work was completed with support by NASA Exoplanet Research Program grant 14-XRP14-2-0090 to the University of Nebraska at Kearney (PI:  AGJ) and National Science Foundation Astronomy and Astrophysics Research Grant AST-1313268 to Wesleyan University (PI: SR).  The authors thank C.~Huang, M.~Swain, and P.~McQueen for many helpful discussions.  We also thank P.~Arras and C.~Duncan for providing feedback on the manuscript.  Finally, we thank the anonymous referee for their insight and several helpful comments.  The Hobby-Eberly Telescope is a joint project of the University of Texas at Austin, the Pennsylvania State University, Stanford University, Ludwig-Maximilians-Universit\"{a}t M\"{u}nchen, and Georg-August-Universit\"{a}t G\"{o}ttingen and is named in honor of its principal benefactors, William P.~Hobby and Robert E.~Eberly.  This work made use of IDL, the Interactive Data Language\footnote{\url{http://www.harrisgeospatial.com/ProductsandTechnology/Software/IDL.aspx}}; IRAF, the Image Reduction and Analysis Facility\footnote{\url{http://iraf.noao.edu/}}; the SIMBAD Database\footnote{\url{http://simbad.u-strasbg.fr/simbad/}}; the Exoplanet Data Explorer\footnote{\url{http://exoplanets.org/}}; and the Exoplanet Transit Database\footnote{\url{http://var2.astro.cz/ETD/index.php}}.


\bibliography{refs}{}
\bibliographystyle{aasjournal}

\end{document}

%% file: macros.tex
\newcommand{\msun}{{~{\rm M}_\odot}}

\newcommand{\kmpers}{{{\rm \; km\;s}^{-1}}}
\newcommand{\ppm}{{\rm \; ppm}}

\newcommand{\eqw}{{W_{\lambda}}}